# Visible Light Optical Data Centre Links

Osama Zwaid Alsulami[1], Mohamed O. I. Musa[1], Mohammed T. Alresheedi[2] and Jaafar M. H. Elmirghani[1]
[1]School of Electronic and Electrical Engineering, University of Leeds, LS2 9JT, United Kingdom
[2]Department of Electrical Engineering, King Saud University, Riyadh, Kingdom of Saudi Arabia
**ml15ozma@leeds.ac.uk, m.musa@leeds.ac.uk, malresheedi@ksu.edu.sa, j.m.h.elmirghani@leeds.ac.uk**

**ABSTRACT**
Providing high data rates is one of the big concerns in visible light communication (VLC) systems. This paper introduces a data centre design that use a VLC system for downlink communication. In this work, RYGB laser diodes (LD) are used as transmitters to obtain a high modulation bandwidth. Three types of receivers, wide field of view receiver (WFOVR), 3 branches angle diversity receiver (ADR) and 50 pixels imaging receiver (ImR) are used to examine delay spread and SNR. The proposed system achieved data rates up to 14.2 Gbps using simple on-off-keying (OOK) modulation.

**Keywords**: VLC, laser diode, data centre, WFOVR, ADR, ImR, SNR, data rate.

## 1. INTRODUCTION

Due to the energy efficiency, scalability and flexibility of visible light communication (VLC) systems, it can be used in next generation data centres replacing hundreds of metres of cables or optical fibre. VLC systems have been shown in many studies to be able to transmit video, data and voice content at data rates up to 20 Gbps in indoor settings [1]-[5]. In addition, due to the growth in demand for high data rates, VLC and in general optical wireless (OW) have received wide interest by researchers for transferring data [6]-[10]. The RF spectrum is becoming scarce, where the demand for higher user data rates continues to grow. Achieving very high data rates beyond 10 Gbps and into the Tbps region using the bandwidth available for radio systems is challenging. According to Cisco, mobile Internet traffic over this half of the decade (2016-2021) is expected to increase by 27 times [11]. Given this expectation of dramatically growing demand for data rates, the quest is already underway for alternative spectrum bands beyond radio waves [12]-[14]. The latter are bands currently used and planned for near future systems, such as beyond 5G cellular systems. VLC and optical wireless (OW) systems can provide a license free bandwidth, high security and low-cost alternative to the RF system [15]-[18]. However, they have some limitations such as the absence of line-of-sight (LOS) components in links which reduces the system performance significantly. In addition, inter-symbol interference (ISI) which is caused by the multi-path propagation can affect the system's performance.

In this paper, a data centre based on VLC system as downlink transmitter is introduced. Red, yellow, green, and blue (RYGB) Laser Diodes (LDs) and three types of receivers are used and the developed model considers the effects of multi-path propagation. The RYGB LDs can provide white colour which can be used for indoor illumination as stated in [19] and are used as transmitters to provide a high modulation bandwidth. This paper is organised as follows: The data centre configuration is described in Section 2 and the optical receiver design is presented in Section 3. Section 4 introduces the optical transmitter design and Section 5 shows the simulation results, and the conclusions are provided in Section 6.

## 2. DATA CENTRE CONFIGURATION

As shown in Figure 1a, the data centre pod dimensions (length × width × height) are assumed to be 8 m × 8 m × 3 m in the evaluation. The proposed data centre pod consists of three rows of racks and each row contains 10 racks [20]-[23]. The red rack in each row (see Figure 1a) is used for routing which is the communication coordinator between each row of racks with the internet. The dimensions of each rack are shown in Figure 1b. One meter or more has been set between rows as well as between each row and the walls for ventilation. Reflections up to second order were considered in this work due to the fact that third and higher order reflections have no impact on the received optical power [24], [25]. A ray-tracing algorithm was utilised for modelling reflections from the ceiling, walls and the floor in the room. Thus, each surface in the room was divided into small equal areas $dA$ with a reflection coefficient of $\rho$. The author in [25] showed that, plaster walls reflect light rays close to a Lambertian pattern. Therefore, each surface in the room such as ceiling, walls and floor was modelled as a Lambertian reflector with reflection's coefficient equal to 0.8 for ceiling and walls and 0.3 for the floor [26], [27]. Each element in each surface acts as a small emitter that reflects the received ray in the form of a Lambertian pattern with $n$ (emission order of the Lambertain pattern) equal to 1. The area of the surface elements can play a significant role in the time resolution of the results. When the surface elements is very small, a higher resolution is gained at thecost of longer simulation time. Therefore, 5 cm × 5 cm was chosen as an area of the surface element for the first order reflection, while 20 cm × 20 cm was chosen as an area of the surface element for the second order reflection to keep the computation time of the simulation within a reasonable time [3], [28]. The communication floor (CF) is set at 0.25 m above the floor as shown in Figure 1a which means all VLC is done above the CF. Under the CF rows or racks can be connected through fibre optic cable.

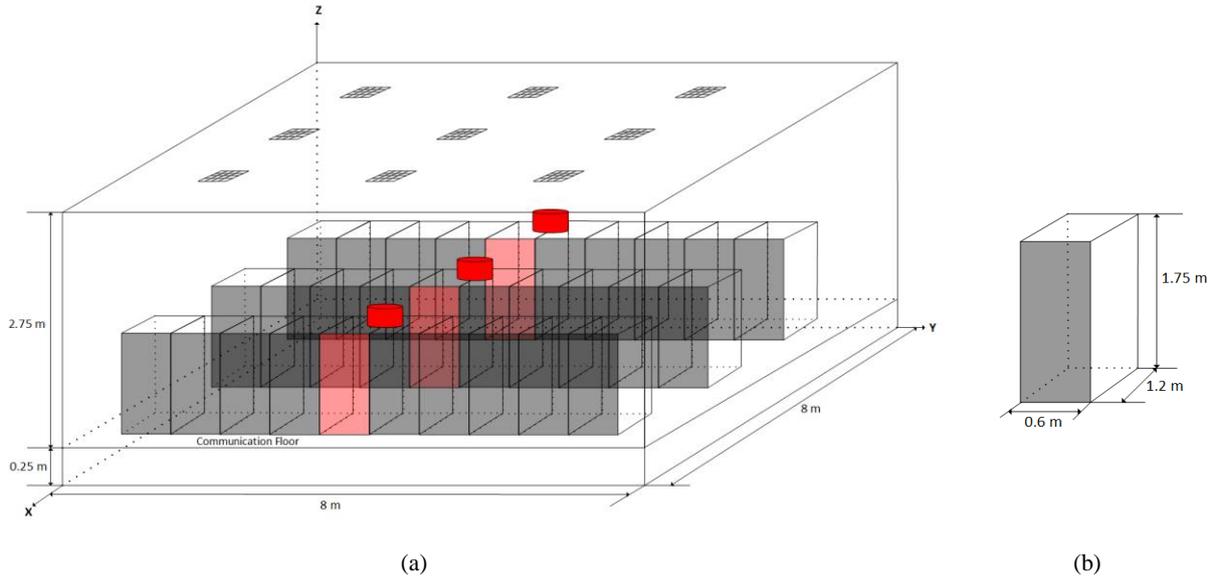

(a)          (b)

Figure 1: (a) Data Centre Configuration, (b) Rack diminutions

## 3. OPTICAL RECEIVER DESIGN

In this work, each row has an optical receiver placed on the top middle of each row (see Figure 1a). Three types of optical receiver have been considered. The first one is a wide field of view receiver (WFOVR) that uses 70° as field of view (FOV) with area equal to 4 $mm^2$ and responsivity equal to 0.4 A/W. The second receiver is an angle diversity receiver (ADR) that has 3 branches of photodetectors with narrow FOVs. The ADR was used to directionally collect signals and reduce the inter-symbol interference (ISI). Each photodetector is oriented to a different direction to cover different transmitters in the data centre ceiling based on the Azimuth ($Az$) and Elevation ($El$) angles. The $El$ angles of the three detectors are set as follow: two detectors are set at 25°, while the detector that faces upwards is positioned at 90°, whereas, the $Az$ angles of the detectors are 0°, 90°and 270°. The FOV of these detectors is set at 20°. In addition, the area of each photodetector is chosen equal to 4 $mm^2$ with responsivity equal to 0.4 A/W. Each receiver of each row covers the three light units placed 1 m directly above the row. The imaging receiver (ImR) is the last type evaluated in this work to collect signals and reduce the ISI. It consists of 50 pixels and each pixel has a narrow FOV (17°) with area equal to 4 $mm^2$ and responsivity equal to 0.4 A/W. Above these pixels, a lens with FOV equal to 65° is used to collect light rays from the three light units that are placed 1 m directly above each row.

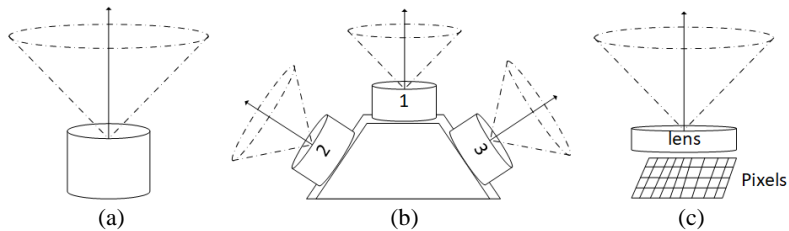

(a)          (b)          (c)

Figure 2: Optical Receiver Design: (a) WFOVR, (b) ADR, (c) ImR.

## 4. OPTICAL TRANSMITTER DESIGN

RYGB LDs light unit are used for illumination and communication in this work. Nine RYGB Light units have been utilised inside the data centre (see Figure 1a). Each unit of these RYGB Light units has 16 wide-semi angles

RYGB LDs. The semi angle of the RYGB LDs was chosen to be 70⁰ to increase the lighting level inside the data centre pod. RYGB LDs light units are located on the ceiling in different locations: (1.8 m, 2 m, 3 m), (1.8 m, 4 m, 3 m), (1.8 m, 6 m, 3 m), (4 m, 2 m, 3 m), (4 m, 4 m, 3 m), (4 m, 6 m, 3 m), (6.2 m, 2 m, 3 m), (6.2 m, 4 m, 3 m) and (6.2 m, 6 m, 3 m) as shown in Figure 1a. The light units positions are chosen in these locations to provide good communication links for each row of racks in the data centre. Therefore, each three RYGB LDs light units placed 1 m above each row of racks are used to transfer data to that row.

## 5. SIMULATION SETUP AND RESULTS

In this work, each three RYGB LDs light units placed directly 1 m above each row of racks carry the same data from the Internet. Thus, each row of racks connects to the Internet separately to provide a high data rate for each row of racks. The receiver which is placed in the top middle of each row receives the visible light signals from the three light units assigned to it and transfers the received data to the router.

Three types of receivers have been examined in this work: WFOVR, ADR and ImR. The SNR should be high to obtain a high data rate in VLC systems to reduce the bit error rate (BER), calculated when using OOK as:

$$BER = Q(\sqrt{SNR}) \qquad (1)$$

where the function $Q(\cdot)$ is the Gaussian function and is expressed as:

$$Q(x) = \frac{1}{2} erfc\left(x/\sqrt{2}\right) \approx \frac{1}{\sqrt{2\pi}} \frac{e^{-(x^2/\sqrt{2})}}{x}. \qquad (2)$$

The SNR in OOK modulation was computed using:

$$SNR = \left(\frac{R(P_{s1}-P_{s0})}{\sigma_t}\right)^2 \qquad (3)$$

where, $R$ is the photodetector responsivity ($R = 0.4\ A/W$), $P_{s1}$ is the power associated with logic 1, $P_{s0}$ is the power associated with logic 0 and $\sigma_t$ is the total noise related to the received signal which can be calculated as:

$$\sigma_t = \sqrt{\sigma_{pr}^2 + \sigma_{bn}^2 + \sigma_{sig}^2} \qquad (4)$$

where, $\sigma_{pr}$ is the preamplifier noise, $\sigma_{bn}$ is the background shot noise and $\sigma_{sig}$ is the shot noise associated with the received signal. The calculation of the background and received signal noises can be found in [29], [30], while the preamplifier noise can be calculated based on the receiver used [31]. Two techniques are used to combine the signals received by the elements of the ADR or received by the pixels of the imaging receiver. Selection combining (SC) involves the selection of a detector with the highest SNR, which is given by:

$$SNR_{SC} = max_k \left(\frac{R(P_{s1}-P_{s0})}{\sigma_t}\right)^2_k, \qquad 1 \leq k \leq J \qquad (5)$$

where, $J$ is the total number of detectors or pixels used. Maximum ratio combining (MRC) is the second combining method, and it combines all the outputs from all the detectors or pixels. The MRC's SNR was calculated using:

$$SNR_{MRC} = \sum_{k=1}^{J} \left(\frac{R(P_{s1}-P_{s0})}{\sigma_t}\right)^2, \qquad 1 \leq k \leq J \qquad . \qquad (6)$$

The imaging receiver employs a lens as shown in Figure 2c that has a transmission factor which depends on the incidence angle ($\Upsilon$) and is given by [32]:

$$Tc(\Upsilon) = -0.1982\Upsilon^2 + 0.0425\Upsilon + 0.8778$$

Figure 3 shows the delay spread and SNR of the three receiver types. The channel bandwidth of the WFOVR is limited to 200 MHz for this setting, due to its delay spread. Thus, WFOVR cannot support high data rates above 285 Mbps. The ADR and the ImR, however, experience more than 3 orders of magnitude lower delay spread, as shown in Figure. 3b, resulting in higher channel bandwidth, and enabling them to support data rates up to 14.2 Gbps for each row inside the data centre. Figure 3b also shows that the delay spread for the imaging receiver is 10 times less than that of the angle diversity receiver.

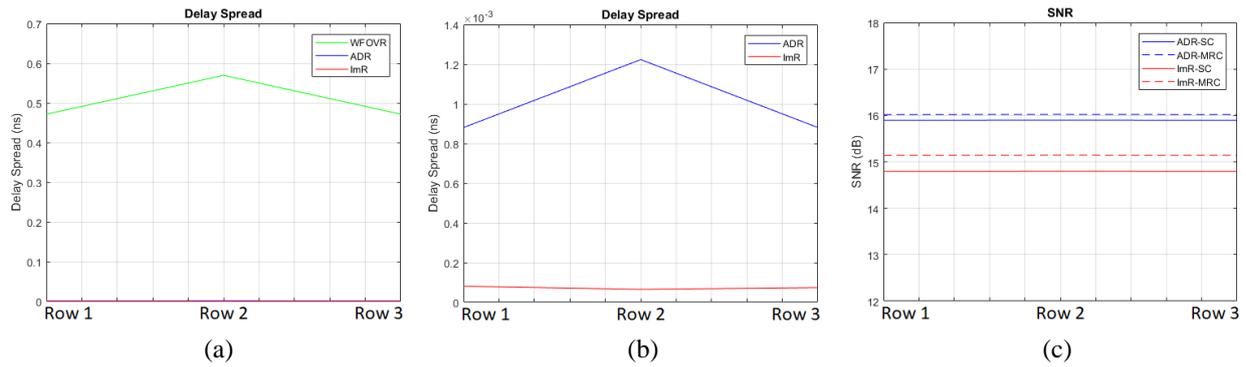

Figure 3: (a) Delay Spread of WFOR, ADR and ImR, (b) Delay Spread of ADR and ImR, (c) SNR.

Figure 3c compares the Signal to noise ratio (dB) of the imaging and the angle diversity receivers using the two combining techniques for each. The results show a 1dB advantage for the ADR receiver over the imaging receiver because both ADR and ImR can collect signals from just 3 light units. Thus, the received power by the ImR is lower than the received power by the ADR due to the transmission factor of the lens used in the imaging receiver. In both receiver types, the MRC was better in terms of SNR than the selective combiner, however, the difference between the two combiners for the ADR receiver is not as high compared to imaging receiver. This is due to the MRC's ability to maximise the SNR by combining all the outputs from all the detectors.

## 6. CONCLUSIONS

A data centre design that use a VLC system for downlink communication was proposed in this paper. The data centre pod considered consists of 30 racks divided into three rows. Each row includes 10 racks, with one of them acting as a router. The proposed system was examined by using different types of receivers (wide field of view receiver (WFOVR), angle diversity receiver (ADR) and imaging receiver (ImR)). In addition, a simple modulation technique, OOK, was utilised in this work. The delay spread was reduced more than 30 times when the ADR and ImR were used. Moreover, the ImR reduced the delay spread 10 times compared to the ADR. However, the ADR can improve the SNR 1 dB over the ImR. A data rate up to 14.2 Gbps can be achieved in each row of racks.

## ACKNOWLEDGEMENTS

The authors would like to acknowledge funding from the Engineering and Physical Sciences Research Council (EPSRC) for the TOWS project (EP/S016570/1). All data are provided in full in the results section of this paper.

## REFERENCES

1. A. T. Hussein and J. M. H. Elmirghani, "10 Gbps Mobile Visible Light Communication System Employing Angle Diversity, Imaging Receivers, and Relay Nodes," J. Opt. Commun. Netw., vol. 7, no. 8, p. 718-735, 2015.
2. S. H. Younus and J. M. H. Elmirghani, "WDM for high-speed indoor visible light communication system," in International Conference on Transparent Optical Networks, 2017.
3. A. T. Hussein, M. T. Alresheedi, and J. M. H. Elmirghani, "20 Gb/s Mobile Indoor Visible Light Communication System Employing Beam Steering and Computer Generated Holograms," J. Light. Technol., vol. 33, no. 24, pp. 5242–5260, 2015.
4. A.T. Hussein, M.T. Alresheedi and J.M.H. Elmirghani, "Fast and Efficient Adaptation Techniques for Visible Light Communication Systems," IEEE/OSA Journal of Optical Communications and Networking, vol. 8, No. 6, pp. 382-397, 2016.
5. S.H. Younus, A. A. Al-Hameed, A. T. Hussein, M.T. Alresheedi, and J.M.H. Elmirghani, "Parallel Data Transmission in Indoor Visible Light Communication Systems," IEEE Access, vol. 7, pp. 1126 - 1138, 2019.
6. A. Al-Ghamdi, and J.M.H. Elmirghani, "Optimisation of a PFDR antenna in a fully diffuse OW system influenced by background noise and multipath propagation," IEEE Transactions on Communication, vol. 51, No. 12, pp. 2103-2114, 2003.
7. A. G. Al-Ghamdi and J. M. H. Elmirghani, "Characterization of mobile spot diffusing optical wireless systems with receiver diversity," ICC'04 IEEE International Conference on Communications, Vol. 1, pp. 133-138, Paris, 20-24 June 2004.
8. A. Al-Ghamdi, and J.M.H. Elmirghani, "Performance evaluation of a triangular pyramidal fly-eye diversity detector for optical wireless communications," IEEE Communications Magazine, vol. 41, No. 3, pp. 80-86, 2003.
9. K.L. Sterckx, J.M.H. Elmirghani and R.A. Cryan, "Sensitivity assessment of a three-segment pyrimadal fly-eye detector in a semi-disperse optical wireless communication link," IEE Proceedings Optoelectronics, vol. 147, No. 4, pp. 286-294, 2000.
10. F.E. Alsaadi, M.A. Alhartomi and J.M.H. Elmirghani, "Fast and Efficient Adaptation Algorithms for Multi-gigabit Wireless Infrared Systems," IEEE/OSA Journal of Lightwave Technology, vol. 31, No. 23, pp. 3735-3751, 2013.

11. C. Mobile, Cisco Visual Networking Index: Global Mobile Data Traffic Forecast Update, 2016-2021 White Paper. 2017.
12. F.E. Alsaadi, and J.M.H. Elmirghani, "Performance evaluation of 2.5 Gbit/s and 5 Gbit/s optical wireless systems employing a two dimensional adaptive beam clustering method and imaging diversity detection," IEEE Journal on Selected Areas in Communications, vol. 27, No. 8, pp. 1507-1519, 2009.
13. M.T. Alresheedi, and J.M.H. Elmirghani, "Hologram selection in realistic indoor optical wireless systems with angle diversity receivers," IEEE/OSA Journal of Optical Communications and Networking, vol. 7, No. 8, pp. 797-813, 2015.
14. M. Alresheedi, and J.M.H. Elmirghani, "10 Gb/s indoor optical wireless systems employing beam delay, power, and angle adaptation methods with imaging detection," IEEE/OSA Journal of Lightwave Technology, vol. 30, pp.1843-1856, 2012.
15. K.L. Sterckx, J.M.H. Elmirghani and R.A. Cryan, "Pyramidal Fly-Eye Detection Antenna for Optical Wireless Systems," Digest IEE Colloq. on Optical Wireless Communications, Digest No. 1999, pp. 5/1-5/6, London, June 1999.
16. F.E. Alsaadi, M.N. Esfahani, and J.M.H. Elmirghani, "Adaptive mobile optical wireless systems employing a beam clustering method, diversity detection and relay nodes," IEEE Transactions on Communications, vol. 58, No. 3, pp. 869-879, 2010.
17. F.E. Alsaadi and J.M.H. Elmirghani, "Adaptive mobile line strip multibeam MC-CDMA optical wireless system employing imaging detection in a real indoor environment," IEEE Journal on Selected Areas in Communications, vol. 27, No. 9, pp. 1663-1675, 2009.
18. M. Alresheedi and J.M.H. Elmirghani, "Performance Evaluation of 5 Gbit/s and 10 Gbit/s Mobile Optical Wireless Systems Employing Beam Angle and Power Adaptation with Diversity Receivers," IEEE Journal on Selected Areas in Communications, vol. 29, No. 6, pp. 1328 - 1340, 2011.
19. A. Neumann, J. J. Wierer, W. Davis, Y. Ohno, S. R. J. Brueck, and J. Y. Tsao, "Four-color laser white illuminant demonstrating high color-rendering quality," Opt. Express, vol. 19, no. S4, p. A982, 2011.
20. X. Dong, T.E.H. El-Gorashi, and J.M.H. Elmirghani, "Green IP over WDM Networks with Data Centres," IEEE/OSA Journal of Lightwave Technology, vol. 27, No. 12, pp. 1861 - 1880, 2011.
21. L. Nonde, T.E.H. El-Gorashi, and J.M.H. Elmirghani, "Energy Efficient Virtual Network Embedding for Cloud Networks," IEEE/OSA Journal of Lightwave Technology, vol. 33, No. 9, pp. 1828-1849, 2015.
22. H.M.M. Ali, A.Q. Lawey, T.E.H. El-Gorashi, and J.M.H. Elmirghani, "Future Energy Efficient Data Centers With Disaggregated Servers," IEEE/OSA Journal of Lightwave Technology, vol. 35, No. 24, pp. 5361 – 5380, 2017.
23. A.M. Al-Salim, A. Lawey, T.E.H. El-Gorashi, and J.M.H. Elmirghani, "Energy Efficient Big Data Networks: Impact of Volume and Variety," IEEE Transactions on Network and Service Management, vol. 15, No. 1, pp. 458 - 474, 2018.
24. J. R. Barry, J. M. Kahn, W. J. Krause, E. A. Lee, and D. G. Messerschmitt, "Simulation of Multipath Impulse Response for Indoor Wireless Optical Channels," IEEE J. Sel. Areas Commun., 1993.
25. F. R. Gfeller and U. Bapst, "Wireless In-House Data Communication via Diffuse Infrared Radiation," Proc. IEEE, vol. 67, no. 11, pp. 1474–1486, 1979.
26. F.E. Alsaadi and J.M.H. Elmirghani, "Mobile Multi-gigabit Indoor Optical Wireless Systems Employing Multibeam Power Adaptation and Imaging Diversity Receivers," IEEE/OSA Journal of Optical Communications and Networking, vol. 3, No. 1, pp. 27-39, 2011.
27. M.T. Alresheedi, A. T. Hussein, and J.M.H. Elmirghani, "Uplink Design in VLC Systems with IR Sources and Beam Steering," IET Communications, vol. 11, No. 3, pp. 311-317, 2017.
28. A. G. Al-Ghamdi and J. M. H. Elmirghani, "Line Strip Spot-Diffusing Transmitter Configuration for Optical Wireless Systems Influenced by Background Noise and Multipath Dispersion," IEEE Trans. Commun., vol. 52, No. 1, pp. 37-45, 2004.
29. J. M. Kahn, "Wireless infrared communications," Proc. IEEE, vol. 85, no. 2, pp. 265–298, 1997.
30. F. E. Alsaadi and J. M. H. Elmirghani, "High-speed spot diffusing mobile optical wireless system employing beam angle and power adaptation and imaging receivers," J. Light. Technol., vol. 28, no. 16, pp. 2191–2206, 2010.
31. E. Kimber, B. Patel, I. Hardcastle, and A. Hadjifotiou, "High performance 10 Gbit/s pin-FET optical receiver," Electron. Lett., vol. 28, no. 2, pp. 120–122, 1992.
32. P. Djahani and J. M. Kahn, "Analysis of infrared wireless links employing multibeam transmitters and imaging diversity receivers," IEEE Trans. Commun., vol. 48, no. 12, pp. 2077–2088, 2000.